\input phyzzx.tex
\tolerance=1000
\voffset=-0.0cm
\hoffset=0.7cm
\sequentialequations
\def\rl{\rightline}

\def\t1{{\tilde 1}}

\def\t{\theta}

\REF{\EDI}{E. Halyo, [arXiv:1502.01979]; [arXiv:1503.07808].}
\REF{\CAR}{J. L. Cardy, Nucl. Phys. {\bf B463} (1986) 435.}
\REF{\WAL}{R. M. Wald, Phys. Rev. {\bf D48} (1993) 3427, [arXiv:gr-gc/9307038]; V. Iyer and R. M. Wald, Phys. Rev. {\bf D50} (1994) 846, [arXiv:gr-qc/9403028]; Phys. Rev. {\bf D52} (1995) 4430, [arXiv:gr-qc/9503052].}
\REF{\ENT}{T. Azeyanagi, T. Nishioka and T. Takayanagi, Phys. Rev. {\bf D77} (2008) 064005, [arXiv:0710.2956].}
\REF{\SEN}{A. Sen, Int. Journ. Mode. Phys. {\bf A24} (2009) 4225, [arXiv:0809.3304]; I. Mandal and A. Sen, Quant. Grav. {\bf 27} (2010) 214003, [arXiv:1008.3801].}
\REF{\NHO}{H. K. Kunduri, J. Lucetti and H. S. Reall, Class. Quant. Grav. {\bf 24} (2007) 4169, [arXiv:0705.4214];
P. Figueras, H. K. Kunduri, J. Lucetti and M. Rangamani, Phys. Rev. {\bf D78} (2008) 044042, [arXiv:0803.2998].}
\REF{\STR}{T. Hartman, K. Murata, T. Nishioka and A. Strominger, JHEP {\bf 0904} (2009) 019, [arXiv:0811.4393].}
\REF{\DAB}{see for example, A. Dabholkar, Lect. Notes Phys. {\bf 851} (2012) 165, [arXiv:1208.4814].}
\REF{\FOR}{A. Sen, JHEP {\bf 0509} (2005) 038, [arXiv: hep-th/0506177]; Gen. Rel. Grav. {\bf 40} (2008) 2249, [arXiv: 0708.1270].}
\REF{\RIN}{M. Parikh and P. Samantray, [arXiv:1211.7370].}
\REF{\BEK}{E. Halyo, JHEP {\bf 1004} (2010) 097, [arXiv:0906.2164].}
\REF{\LEN}{L. Susskind, [arXiv:hep-th/9309145].}
\REF{\EDIW}{E. Halyo. [arXiv:1403.2333]; [arXiv:1406.5763].}
\REF{\SBH}{E. Halyo, A. Rajaraman and L. Susskind, Phys. Lett. {\bf B392} (1997) 319, [arXiv:hep-th/9605112].}
\REF{\EDII}{E. Halyo, Int. Journ. Mod. Phys. {\bf A14} (1999) 3831, [arXiv:hep-th/9610068]; Mod. Phys. Lett. {\bf A13} (1998), [arXiv:hep-th/9611175].}
\REF{\DES}{E. Halyo, [arXiv:hep-th/0107169].}
\REF{\UNI}{E. Halyo, JHEP {\bf 0112} (2001) 005, [arXiv:hep-th/0108167]; [arXiv:hep-th/0308166].}
\REF{\TRA}{D. Birmingham, K. S. Gupta and S. Sen, Phys.Lett. {\bf B505} (2001) 191, [arXiv:hep-th/0102051]; 
A. J. M. Medved, D. Martin and M. Visser, Phys.Rev. {\bf D70} (2004) 024009 [arXiv:gr-qc/0403026].} 
\REF{\ADS}{A. Strominger, JHEP {\bf 9901} (1999) 007, [arXiv:hep-th/9809027].}
\REF{\ABH}{T. Hartman and A. Strominger, JHEP {\bf 0904} (2009) 026, [arXiv:0803.3621]; A. Castro, D. Grumiller, F. Larsen and R. McNees, jHEP {\bf 0811} (2008) 052, [arXiv:0809.4264].} 
\REF{\BTZ}{A. Strominger, JHEP {\bf 9802} (1998) 009, [arXiv:hep-th/9712251].}

\singlespace
\rl{SU-ITP-15/09}
\pagenumber=0
\normalspace
\medskip
\bigskip
\titlestyle{\bf{Extremal Black Hole Entropy from Horizon Conformal Field Theories}}
\smallskip
\author{ Edi Halyo{\footnote*{e--mail address: halyo@stanford.edu}}}
\smallskip
\centerline {Department of Physics} 
\centerline{Stanford University} 
\centerline {Stanford, CA 94305}
\smallskip
\vskip 2 cm
\titlestyle{\bf ABSTRACT}

We show that the entropy of extremal $D=4$ Reissner--Nordstrom black holes can be computed from horizon CFTs with central charges and conformal weights fixed by the dimensionless Rindler energy. This is possible in the simultaneous extremal and near horizon limit of the black hole which takes the geometry to an $AdS_2$ Rindler space with finite temperature. The CFT description of dilatonic $AdS_2$ black holes, obtained from extremal ones by dimensional reduction, lead to exactly the same CFT states.

\singlespace
\vskip 0.5cm
\endpage
\normalspace

\centerline{\bf 1. Introduction}
\medskip

Recently, it was shown that all nonextreme black holes can be described by the states of chiral $D=2$ conformal field theories 
(CFTs) that live in the very near
horizon region[\EDI]. The central charges of the CFTs and the conformal weights of the states that correspond to black holes are determined by the dimensionless Rindler energy $E_R$ as $c/12=L_0=E_R$. The black hole entropy is simply the entropy of the CFT state which is given by the Cardy formula[\CAR]. This can be generalized to all theories of gravity in which case the Cardy formula correctly reproduces the Wald entropy[\WAL] of nonextreme black holes. We stress that the
nonextremality of the black hole, i.e. the fact that its near horizon region is Rindler space is crucial for the description. Obviously, $E_R$ can be defined only in Rindler space. Moreover, the central charges and conformal weights of the black hole states are fixed by requiring that the horizon CFT temperature matches the dimensionless Rindler temperature of the black hole.

At first thought, it seems that the above description of nonextreme black holes in terms of horizon CFTs cannot possibly apply to extremal black holes with nonzero entropy but vanishing temperature.
Taking first the extremal and then the near horizon limit of charged black holes gives rise to an $AdS_2$ factor in the geometry
(times an internal compact space) or equivalently to an $SO(2,1)$ factor in their isometry groups. This is certainly true for $D=4$ Riessner--Nordstrom (RN) black holes which we consider explicitly in this paper. Just like extremal black holes, $AdS_2$ has a vanishing temperature but nonzero (entanglement) entropy[\ENT]. 
In these cases, since the near horizon region is no longer Rindler space the horizon CFT description does not seem to make sense. 

On the other hand, we can take the 
near horizon and extremal limits of the $D=4$ RN black hole geometry simultaneously. This limit gives rise to an $AdS_2$ Rindler space (times an internal compact space) which is not extremal, i.e. with an acceleration horizon and a finite temperature[\SEN].
This surprising result is due to the peculiar nature of the limit in which the original time parameter is taken to infinity
simultaneously with the near horizon limit.
We expect the same (type of) limit to lead to Rindler $AdS_2$ spaces for all extremal charged black holes due to the common $AdS_2$ factor in their near horizon geometries[\NHO].
Since $AdS_2$ Rindler space is nonextreme, we can describe it by the horizon CFT mentioned above. We show that 
the horizon CFT state that describes the $AdS_2$ Rindler space has exactly the correct extremal black hole entropy. Thus, horizon CFTs seem to provide a unified description of both extremal and nonextreme black holes.

Extremal RN black holes have been described in terms of CFTs that live on their horizons in ref. [\STR].
We find that in our description, the central charge (conformal weight) of the black hole state is $Q$ times smaller (larger) 
than those in ref. [\STR]. This is because we identify the CFT temperature with the dimensionless Rindler temperature , i.e $T_{CFT}=T_R=1/2 \pi$ rather than with the Hawking temperature, i.e. $T_{CFT}=T_H=1/ 2 \pi Q$.

This paper is organized as follows. In the next section, we briefly review $D=4$ RN black holes and the simultaneous
extremal and near horizon limit of their metrics. In section 3, we obtain the entropy of these black holes from horizon CFTs by computing the dimensionless Rindler energy $E_R$. In section 4, we show that dilatonic $AdS_2$ black holes with nonzero electric fields, which are obtained from the dimensional reduction of extremal $D=4$ RN black holes, are also described by CFT states with exactly the same central charges and conformal weights as those arising from the horizon CFTs. Section 5 contains a discussion of our results and our conclusions.

\bigskip
\centerline{\bf 2. Near Horizon Limit of Extremal Charged Black Holes}
\medskip
In this section, we review the properties of $D=4$ RN black holes and the simultaneous near horizon and extremal limit
of their geometry. In this special limit, the geometry becomes an $AdS_2$ Rindler space ($\times S^2$) which has, unlike the original
extremal black hole, a nonzero temperature. Therefore, we can use methods of computing nonextreme black hole entropy in this case. 
The near horizon geometries of all extremal charged black holes contain
an $AdS_2$ Rindler factor or an $SO(2,1)$ isometry. Thus, even though in this paper we
explicitly consider only extremal $D=4$ RN black holes, we expect our results to hold for all extremal charged black holes.

Consider the $D=4$  RN black hole with the metric (we set $G=1$ in the following)[\DAB]
$$ds^2=-\left(1-{{2M} \over r}+ {Q^2 \over r^2}\right)dt^2+\left(1-{{2M} \over r}+ {Q^2 \over r^2}\right)^{-1} dr^2+r^2 d\Omega^2
\quad, \eqno(1)$$
with the electromagnetic field strength $F_{tr}=Q/r^2$. The outer and inner horizons, $r_+$ and $r_-$ respectively,
are given by
$$r_{\pm}=M\pm\sqrt{M^2-Q^2} \quad. \eqno(2)$$
The Bekenstein-Hawking entropy is 
$$S_{BH}=\pi r_+^2=\pi(M+\sqrt{M^2-Q^2})^2 \quad, \eqno(3)$$
whereas the Hawking temperature of the black hole is given by
$$T_H={{r_+-r_-} \over {4 \pi r_+^2}}={{\sqrt{M^2-Q^2}} \over {2\pi(2M(M+\sqrt{M^2-Q^2})-Q^2)}} \quad. \eqno(4)$$
In order to satisfy the cosmic censorship hypothesis the mass must satisfy the bound $M \geq Q$.
The extremal limit of the black hole is given by $M \to Q$ or $r_+ \to r_-$. The extremal metric is given by
$$ds^2=-\left(1-{Q \over r}\right)^2dt^2+\left(1-{Q \over r}\right)^{-2}dr^2+r^2d\Omega^2 \quad, \eqno(5)$$
where the horizon is at $r_+=r_-=Q$. $T_H$ vanishes for the extremal black hole whereas the entropy, $S=\pi Q^2$, is finite.
It is well--known that the near horizon limit, with $r_+ \to 0$, of eq. (5) is $AdS_2 \times S^2$ (or the Robinson--Bertotti space) with the metric in isotropic coordinates, $\rho=Q^2/r$,
$$ds^2=\left(-{\rho^2 \over Q^2}dt^2+{Q^2 \over \rho^2}d\rho^2 \right)+Q^2 d\Omega^2 \quad. \eqno(6)$$
We see that both $AdS_2$ and $S^2$ have a radius of $Q$. 
We stress that $AdS_2 \times S^2$ is obtained by taking first the extremal and then the near horizon limit. The temperature of the $AdS_2$ space vanishes just like that of the extreme black hole.
Thus, the nonzero entropy of the extreme black hole has to correspond to that of $AdS_2$. It has been shown that this entropy is
precisely the entanglement entropy of $AdS_2$ arising from the entanglement between the its two disjoint boundaries[\ENT].

One can also take the extremal and near horizon limits simultaneously. The metric in eq. (1) can also be written as
$$ds^2=-\left(1-{r_+ \over r}\right)\left(1-{r_- \over r}\right)dt^2+\left(1-{r_+ \over r}\right)^{-1}
\left(1-{r_- \over r}\right)^{-1}dr^2 +r^2 d\Omega^2 \quad, \eqno(7)$$
where $r_{\pm}$ are defined in eq. (2). We now take the extremal limit $r_+ \to r_-$ keeping the two dimensionless quantities[\SEN]
$$\sigma={(2r-r_+-r_-) \over (r_+-r_-)} \qquad \tau={{(r_+-r_-)t} \over {2 r_+^2}} \quad, \eqno(8)$$
fixed. We note that keeping $\sigma$ fixed when $r_+ \to r_-$ also means $r \to r_+$. Thus the extremal and near horizon limits
are taken simultaneously. On the other hand, keeping $\tau$ fixed in the extremal limit means that we are taking the $t \to \infty$
limit. In this limit, the metric in eq. (7) becomes
$$ds^2=Q^2 \left(-(\sigma^2-1)d\tau^2+{d\sigma^2 \over{\sigma^2-1}}\right)+Q^2 d\Omega^2 \quad. \eqno(9)$$
since in the extremal limit $r_+=M=Q$.
The metric above describes $AdS_2$ Rindler space ($\times S^2$ of radius $Q$) with an acceleration horizon at $\sigma=1$. This can be easily seen by using the coordinate transformation $\rho=\sqrt{\sigma^2-1}$ which takes the metric in eq. (9) to[\RIN]
$$ds^2=Q^2\left({\rho^2}d\tau^2+{d\rho^2 \over {1+\rho^2}}\right)+r_+^2 d\Omega^2 \quad. \eqno(10)$$
For $\rho<<1$ the first two terms describe Rindler space (with the horizon at $\rho=0$) whereas for $\rho>>1$ they become the 
Poincare patch of $AdS_2$. 
In refs. [\SEN], the macroscopic entropy of the space given by the metric in eq. (9) was calculated by using the entropy function formalism[\FOR] (which works for all spaces with an $AdS_2$ factor) and shown to match the entropy of the extremal $D=4$ RN black hole. It is interesting to note that $AdS$ Rindler space (in any dimension) saturates the Bekenstein bound on the entropy of the boundary CFT[\BEK].


\bigskip
\centerline{\bf 3. Extremal Charged Black Hole Entropy from Horizon CFTs}
\medskip

We can compute the entropy of the $D=4$ extremal RN black hole by starting with a nonextreme RN black hole and then taking the extremal limit. The nonextreme black hole can be described by a horizon CFT state. Therefore, the extremal black hole can be described by the extremal limit of the horizon CFT which, using the Cardy formula, gives the correct entropy. However, in the extremal limit, the near horizon geometry is $AdS_2$ rather than Rindler space, so the meaning of quantities such as the dimensionless Rindler energy (which gives the entropy) is not clear. In addition, it seems strange to describe a black hole with vanishing temperature in terms of a CFT at a finite one.

On the other hand, we can take the limit in eq. (8) and consider the metric in eq. (9).
Unlike the extremal charged black hole we started with, the space described by the metric in eq. (9) has a nonextreme horizon at $\sigma=1$. The near horizon region of this metric is Rindler space just like those of nonextreme black holes. 
Thus, we can use the same methods used for computing nonextreme black hole entropy in the present case even though the original black hole is extremal.
For example, we can calculate the entropy of $AdS_2$ Rindler space ($\times S^2$) by computing the dimensionless Rindler energy,
$E_R$, of the metric in eq. (9) and using $S_{BH}=2 \pi E_R$[\LEN,\EDIW]. This relation was used to compute the entropy of 
a number of nonextreme black holes and de Sitter space[\SBH-\UNI]. 
We then interpret the black hole entropy 
as the entropy of a state with $c/12=L_0=E_R$ in a $D=2$ chiral CFT that lives on the black hole horizon[\EDI]. 

Consider the near horizon region of the metric in eq. (9) where $\sigma=1+y$ and $y<<1$. In terms of the proper distance to the horizon, $\rho=\sqrt{y}$, the metric becomes
$$ds^2=Q^2(-\rho^2 d\tau^2+d\rho^2+d\Omega^2) \quad. \eqno(11)$$
The dimensionless Rindler temperature of this space is $T=1/2\pi$ whereas the Hawking temperature is given by $T_H=1/2\pi Q$.
$E_R$ is the dimensionless Rindler energy that is conjugate to the dimensionless Rindler time $\tau$ in eq. (11).
Using the fact that in the extremal limit $M \to Q$ we obtain $E_R$[\LEN,\EDIW]
$$E_R=\int {dM \over {2 \pi T_H}}={Q^2 \over 2} \quad. \eqno(12)$$
The Bekenstein--Hawking entropy of the black hole is 
$$S_{BH}=2 \pi E_R=\pi Q^2 \quad, \eqno(13)$$
reproducing eq. (3).

For nenextreme black holes, the relation $S=2 \pi E_R$ can be interpreted as the entropy of a $D=2$ chiral CFT state that lives
in the very near horizon region[\EDIW]. The near horizon region of $AdS_2$ Rindler space with metric in eq. (9) is Rindler space given by eq. (11). In the very near horizon region, i.e. near the origin of Rindler space, all dimensionful parameters can be ignored and the physics is described by a CFT. In addition, the directions that are transverse to $\tau$ and $\rho$ decouple[\TRA]
resulting in a $D=2$ CFT. This CFT is chiral since Rindler space has a single $U(1)$ isometry which gets enhanced to a single Virasoro
algebra near the horizon. 
If we demand that the CFT temperature, $T_{CFT}$, defined as
$$T_{CFT}={1 \over \pi} \sqrt{{{6 \Delta} \over c}} \quad, \eqno(14)$$
where $\Delta$ is the conformal weight in Rindler space, matches the dimensionless Rindler temperature, i.e. $T_{CFT}=T_R=1/2 \pi$,
we find that he CFT has a central charge of $c/12=E_R$ whereas the conformal weight of the state that
corresponds to the black hole is $L_0=E_R$. 
Due to the exponential mapping from the Minkowski to the Rindler space, the conformal
weights in Rindler space, $\Delta$ are shifted by $-c/24=E_R/2$ with respect to those in Minkowski space, $L_0$[\EDI]. Thus, 
$\Delta=L_0-c/24=L_0/2$. The entropy of this CFT state 
is then given by the Cardy formula[\CAR]
$$S=2\pi \sqrt{{{c \Delta} \over 6}}=2 \pi E_R=\pi Q^2 \quad, \eqno(15)$$
reproducing eq. (3) for the extremal RN black hole entropy.

We conclude that in the simultaneous extremal and near horizon limit, a charged black hole is described by a state of a $D=2$ CFT that lives in the very near horizon region with[\EDI]
$${c \over 12}=L_0=E_R={Q^2 \over 2} \quad, \eqno(16)$$
and a shift of $-E_R/2$ in the conformal weights.

Extremal RN black holes were described as slightly different CFT states in ref. [\STR]. According to ref. [\STR], the black hole corresponds to a CFT state with $c=6Q^3$ and $T_{CFT}=1/2 \pi Q$ which means that $L_0=Q/2$. We observe that $c$ ($L_0$) is $Q$ times larger (smaller) 
than our values in eq. (16). This is because we equated the CFT temperature to the dimensionless Rindler temperature , i.e $T_{CFT}=T_R=1/2 \pi$ rather than to the Hawking temperature, i.e. $T_{CFT}=T_H=1/ 2 \pi Q$. The First Law of Thermodynamics
relates $T_R$ and $T_H$ to the dimensionless Rindler energy $E_R$ and the extremal mass $Q$ respectively.
If we identified $T_{CFT}=T_H$ above, we would obtain CFT states that agree with those of ref. [\STR].


\bigskip
\centerline{\bf 4. $D=2$ Dilatonic Black Holes and Horizon CFTs}
\medskip

In the previous section we found that charged extremal black holes can be described by states in a $D=2$ CFT with $c/12=L_0=Q^2/2$.
We obtained this result by finding $E_R$ and interpreting the black hole entropy $S=2 \pi E_R$ as that of the CFT states. On the other hand, the extremal limit of the $D=4$ RN black hole, in the near horizon limit can be dimensionally reduced to an $AdS_2$ black hole with fixed dilaton and electric field. This solution can be described as a particular $D=2$ CFT state.
In this section, we show that this CFT state is exactly the one predicted by the horizon CFT of the previous section. This
provides additional support for our description.

In the near horizon limit, the $D=4$ RN black hole can be dimensionally reduced to an $AdS_2$ space with a
constant dilaton and electric field[\ADS].
This is a solution to $D=2$ Maxwell--dilaton gravity with the action[\ABH]
$$I=\alpha \int d^2x \sqrt{-g} \left[e^{-2\phi}\left(R+{8 \over \ell^2} \right)-{\ell^2 \over 4}F^2 \right] \quad, \eqno(17)$$
where $\alpha=-e^{2\phi}/16\pi G_2$, $\phi$ is the dilaton, $F$ is the $U(1)$ field strength and $G_2$ is the two--dimensional Newton's constant given by 
$$G_2={G_4 \over {4 \pi Q^2}}={1 \over {4 \pi Q^2}} \quad, \eqno(18)$$
due to the spherical reduction over $S^2$ with radius $Q$.
The equations of motion obtained from eq. (17) are solved by[\ABH]
$$R=-{8 \over \ell^2} \qquad F_{\mu\nu}=2E\epsilon_{\mu\nu} \qquad e^{-2\phi}=-{\ell^4 \over 4}E^2 \quad, \eqno(19)$$
where $E$ is the electric field. We see that the solution is an $AdS_2$ space with radius $\ell$ and constant electric 
field and dilaton. Theories of two dimensional gravity can be described as CFTs. 
From the transformations of the stress--energy tensor under diffeomorphisms near the $AdS_2$ boundary and the
$U(1)$ gauge transformation one obtains the central charge of the CFT that describes the solution in eq. (19)[\ABH]
$$c=-24 \alpha e^{-2\phi}={3 \over {2\pi G_2}} \quad. \eqno(20)$$
Using eq. (18) we find that $c/12=Q^2/2$ precisely the result obtained from the horizon CFT.

The action in eq. (17) has also black hole solutions with the metric
$$ds^2=-\left({{r^2} \over {\ell^2}}-a^2\right)dt^2+ \left({{r^2} \over {\ell^2}}-a^2\right)^{-1}dr^2 \quad, \eqno(21)$$
with the same constant electric field and dilaton as in eq. (19).
Here $\ell=Q/4$ and $a^2=16 \pi \ell G_2 M$. These black holes are in a one--to--one correspondence with the $D=4$ RN black holes since they are obtained by dimensional reduction of the latter over $S^2$. 

The solutions in eq. (21) correspond to states of a CFT with the central charge given by eq (20).
In the CFT, the conformal weight is $\Delta=M \ell$. For example, global $AdS_2$ which corresponds to $a^2=-1$ or
$M=-1/16 \pi G_2 \ell$ is described by a state with $\Delta=-1/16 \pi G_2=-c/24$. This is the ground state in the NS sector of a
supersymmetric embedding of the theory in eq. (17)[\BTZ]. The zero mass black hole with $M=a=0$ corresponds to a state with $\Delta=0$ which is the ground state in the Ramond sector. The extremal limit of the charged black hole corresponds to a state with $a=1$ or $r=\ell$, which is $AdS_2$ Rindler space, since when $a=1$ we get
$$M={1 \over {16 \pi G_2 \ell}}= Q  \quad. \eqno(22)$$
Then, the CFT state that corresponds to this particular black hole (which is actually $AdS_2$ Rindler space) has $\Delta=c/24$. Using Cardy's formula in eq. (15) we get the correct black hole entropy in eq. (3). 
We found that the dilatonic $AdS_2$ black hole, that corresponds to the extremal $D=4$ RN black hole, is described by  a state in a $D=2$ CFT with $c/12=L_0=E_R=Q^2/2$. This is precisely the same CFT state in the horizon CFT in eq. (16).
This result constitutes supporting evidence for the horizon CFT description of extremal black holes.

We can also find $E_R$ for the metric in eq. (21) and interpret the resulting entropy as that of a state in a horizon CFT.
The horizon of this metric is at $r_h=a \ell$. Near the horizon, we can write $r=r_h+y$ where $y<<r_h$ and find the metric
in terms of the proper distance to the horizon $\rho=\ell \sqrt{2y/r_h}$,
$$ds^2=-{{r_h^2} \over {\ell^4}} \rho^2 dt^2 + d \rho^2 \quad. \eqno(23)$$
The mass and Hawking temperature of this black hole are given by
$$M={{r_h^2} \over {16 \pi \ell^3 G_2}}  \qquad T_H={r_h \over {2\pi \ell^2}}  \quad. \eqno(24)$$
Here $r_h \geq \ell$ which is equivalent to the condition $M \geq Q$ on the $D=4$ RN black hole mass. 
The dimensionless Rindler energy is 
$$E_R= \int {dM \over {2 \pi T}}={r_h \over {8 \pi \ell G_2}} \quad, \eqno(25)$$
The extremal limit of the charged black hole corresponds to $r_h=\ell$ (or $a=1$) which leads to the extremal mass $M=Q$.
Then, the extremal entropy is given by
$$S=2 \pi E_R={1 \over {4G_2}}=\pi Q^2 \quad. \eqno(26)$$ 
correctly reproducing the entropy in eq. (3).

\endpage

\bigskip
\centerline{\bf 5. Conclusions and Discussion}
\medskip

In this paper, we showed that the entropy of an extremal $D=4$ RN black hole can be obtained from that of a $D=2$ CFT that lives in the very near horizon region. The CFT state that describes the black hole has a central charge and conformal weight given by
$c/12=L_0=E_R=Q^2/2$ where $E_R$ is the dimensionless Rindler energy and $Q$ is the black hole charge. The black hole entropy is then given by the Cardy formula. 

It is known that horizon
CFTs of this type describe nonextreme black holes with a finite temperature. Therefore, it is surprising that the same CFTs also describe extremal black holes with vanishing temperature. This is explained by the simultaneous extremal and near horizon limit in eq. (8). This limit takes the extremal geometry into an $AdS_2$ Rindler space with nonzero temperature which can be described by a horizon CFT. Thus, horizon CFTs give a uniform description of both extremal and nonextreme black holes.

We also showed that the dimensional reduction (over $S^2$) of the near extremal geometry of $D=4$ RN black holes, which leads to $AdS_2$ black holes, is described by exactly the same CFT states. We interpret this result as supporting evidence for our description.
Even though we explicitly considered only $D=4$ RN black holes in this paper, we expect our results to hold for all extremal charged black holes since these all have a near horizon geometry with an $AdS_2$ factor or $SO(2,1)$ isometry.

The entropy of extremal charged black holes can also be computed by using the
entropy function formalism[\FOR]. This method works for all extremal charged black holes since these have an $AdS_2$ Rindler space factor in the limit given by eq. (8). The entropy function ${\cal E}$ evaluated at its extremum is the entropy black hole[\FOR]
$$S={\cal E}=2 \pi \left(e_i {\partial f \over \partial e_i} -f\right) \qquad, \eqno(27)$$
where $e_i$ are the electric fields and $f$ is the Lagrangian density evaluated at the horizon and integrated over the angular directions
$$f=\int d\theta d \phi \sqrt{-g} {\cal L} \qquad \eqno(28)$$
We see that since Wald entropy is given by both $2\pi E_R$ and eq. (27), 
$$E_R=\left(e_i {\partial f \over \partial e_i} -f\right) \qquad, \eqno(29)$$
At first thought, $E_R$ and ${\cal E}$ do not seem to be related except for the fact that both quantities are computed from the near
horizon geometry. It would be nice to understand their relation from first principles.


\bigskip
\centerline{\bf Acknowledgments}

I would like to thank Lenny Susskind for useful discussions and the Stanford Institute for Theoretical Physics for hospitality.

\vfill

\refout

\end
\bye